# SCIENTIFIC REPORTS

**OPEN**

# New Description of Evolution of Magnetic Phases in Artificial Honeycomb Lattice

B. Summers, Y. Chen, A. Dahal & D. K. Singh

Artificial magnetic honeycomb lattice provides a two-dimensional archetypal system to explore novel phenomena of geometrically frustrated magnets. According to theoretical reports, an artificial magnetic honeycomb lattice is expected to exhibit several phase transitions to unique magnetic states as a function of reducing temperature. Experimental investigations of permalloy artificial honeycomb lattice of connected ultra-small elements, $\simeq$ 12 nm, reveal a more complicated behavior. First, upon cooling the sample to intermediate temperature, $T \simeq 175$ K, the system manifests a non-unique state where the long range order co-exists with short-range magnetic charge order and weak spin ice state. Second, at much lower temperature, $T \simeq 6$ K, the long-range spin solid state exhibits a re-entrant behavior. Both observations are in direct contrast to the present understanding of this system. New theoretical approaches are needed to develop a comprehensive formulation of this two dimensional magnet.

Nanostructured prototypes of geometrically frustrated magnet have attracted lot of attentions in recent years[1,2]. Unlike bulk materials that are used for the investigation of individual phenomenon[3], two-dimensional nanostructured magnet not only provides a facile platform to explore many novel magnetic properties in a single disorder free environment but also paves the way for finding new details of known physical principles[4–6]. Artificial magnetic honeycomb lattice is a prominent research venue in this regard[6,7]. Originally conceived to explore the magnetic analogue of ice-rule and associated Dirac's effective monopoles using standard experimental techniques[8–10], such as magnetic force microscopy and X-ray dichroism method, it has become a subject of extensive investigation to find new properties of geometrically frustrated magnets[7,11]. Synergistic efforts of theoretical and experimental researches on artificial magnetic honeycomb lattice have shown that the system undergoes gradual transition from paramagnetic spin gas state to spin ice state to a magnetic charge ordered state, manifested by neighboring pair of vortex loops of opposite chiralities, as a function of reducing temperature[11–13]. The evolution of magnetic phases in artificial magnetic honeycomb lattice is entropy driven[11,12]. Here, a characteristic temperature, comparable to the inter-elemental magnetostatic energy, plays the key role. At much lower temperature, the system is predicted to manifest a long range ordered ground state of spin solid state with zero entropy and magnetization, consisting of an ordered arrangement of vortex loops of opposite chiralities.

We have used small angle neutron scattering (SANS) and polarized reflectometry measurements on a large sample of permalloy ($Ni_{0.81}Fe_{0.19}$) artificial honeycomb lattice of connected ultra-small element ($\simeq$ 12 nm in length and 5 nm in width), see Fig. 1a, to study this problem. Contrary to the existing understanding that the temperature dependent evolution of magnetism results in unique magnetic phases in an artificial honeycomb lattice, new experimental investigations of nanostructured permalloy honeycomb lattice reveal the unusual coexistence of a long-range ordered state with short-range correlated magnetic charge ordered state at intermediate temperature of $T = 175$ K. It suggests that the system exhibits the tendency to develop competing states as the temperature is reduced below the paramagnetic spin gas phase. When temperature is further reduced to $T \simeq 40$ K, the long-range order disappears and only weak remnants of the short-range order, spanning over the lateral size of two honeycombs $\simeq$ 70 nm, survive. The most surprising behavior, however, is observed at a much lower temperature, $T = 6$ K, when the long-range ordered state reappears as the only magnetic phase of the system. Both phenomena, the non-unique manifestation of temperature dependent phases and the re-entrant characteristic of the long-range ordered state in an artificial honeycomb lattice, are highly unexpected. Our experimental observations are schematically described

Department of Physics and Astronomy, University of Missouri, Columbia, MO 65211, USA. B. Summers and Y. Chen contributed equally to this work. Correspondence and requests for materials should be addressed to D.K.S. (email: singhdk@missouri.edu)

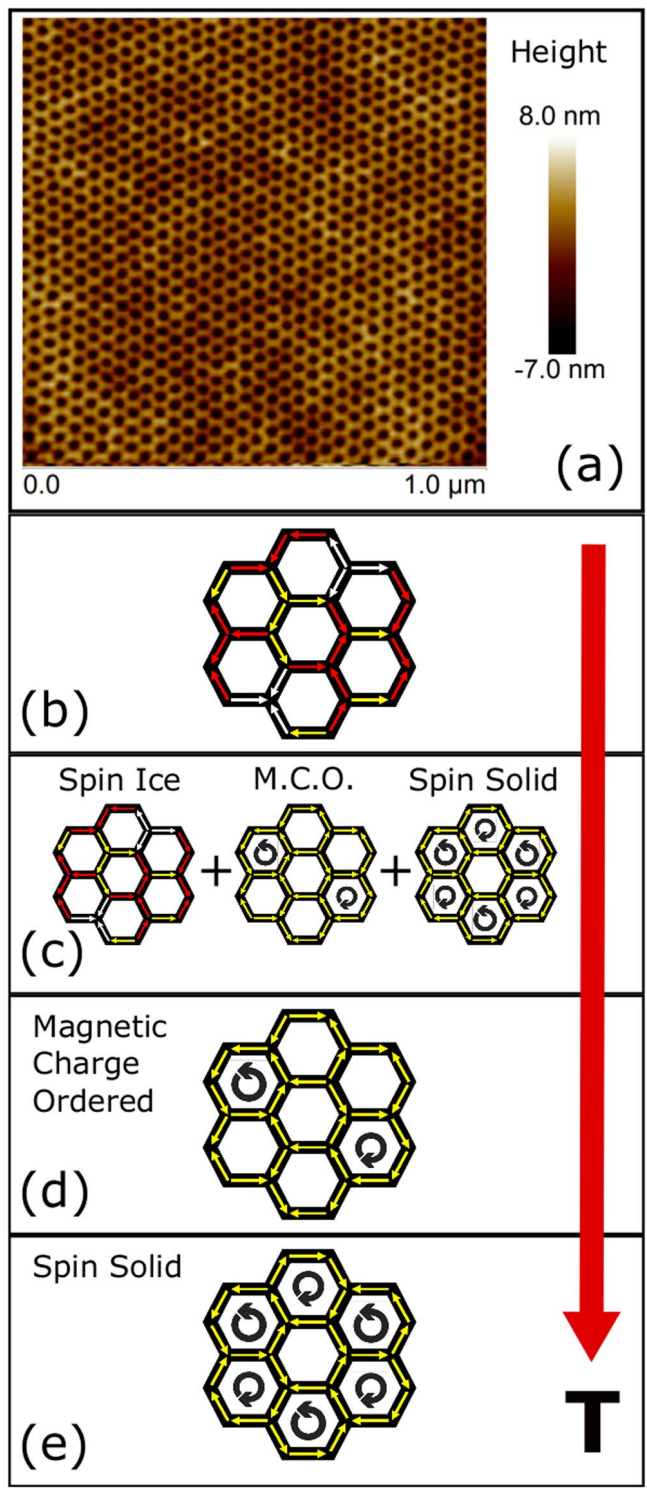

**Figure 1.** Schematic description of temperature dependent magnetic phases. (**a**) Atomic force micrograph of a typical nanostructured artificial honeycomb lattice. (**b–e**) Magnetic phases in artificial honeycomb lattice of ultra-small element. As temperature reduces, the system undergoes through a variety of states: from a paramagnetic spin gas state at high temperature, consisting of random distribution of 2-in & 1-out (or vice-versa) spin ice type arrangements, to a combination of spin ice, magnetic charge ordered state and long range ordered spin solid state in the higher range of intermediate temperature. For further reduction in temperature, the system manifests a short-range ordered state before developing a pure long-range ordered state in $T \rightarrow 0\,K$ limit.

in Fig. 1b–e. The underlying magnetic configuration behind the long-range ordered state is further investigated using polarized neutron reflectometry measurements. The reflectometry data at low temperature is well described by the modeling of spin solid arrangement of correlated moments.

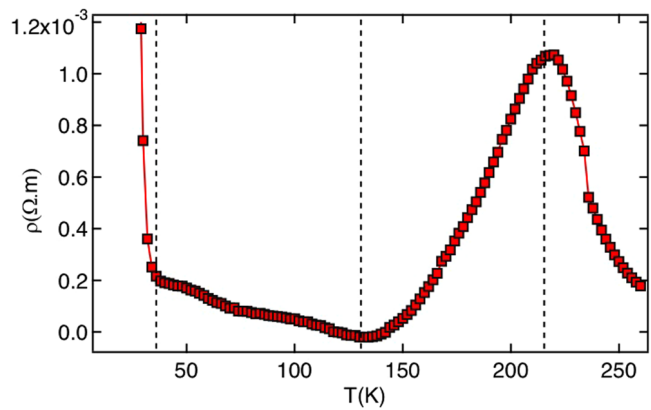

**Figure 2.** Electrical measurement of permalloy artificial honeycomb lattice. Resistivity varies differently in different temperature regimes, as shown by the dashed lines. As explained in the text, the four temperature regimes suggest the existence of different magnetic phases in the system.

For a moderate aspect ratio of structural parameters e.g. length, width and thickness, magnetic moments aligned along the honeycomb elements follow the quasi-ice rule[14]. On a given vertex of the lattice, two possibilities emerge: either all moments point to or away from the vertex, also termed as 'all-in or all-out' configuration, or two moments point to the vertex and one point away from the vertex (or vice versa), termed as 'two-in & one-out' (or vice-versa) configuration[6,15]. These peculiar arrangements of magnetic moments on the vertices of honeycomb lattice and the consequential magnetic transitions are experimentally investigated in both disconnected and connected elemental geometries[7,16,17]. Unlike previous experimental efforts that utilized electron beam lithography technique to fabricate honeycomb samples, we have used diblock template method[18] to create large throughput nanostructured artificial honeycomb lattice of ultra-small element. The typical size of a honeycomb element is 12 nm (length)× 5 nm (width)× 7 nm (thickness). While the ultra-small element geometry ensures that the inter-elemental energy is small enough ($\simeq$15 K) to allow temperature to be a tuning parameter for the investigation of novel magnetic phases, the large sample size makes it suitable for the experimental research using macroscopic probes of small angle neutron scattering (SANS) and polarized reflectometry.

## Results

**Electrical measurements of artificial honeycomb lattice.** Fabrication of nanostructured artificial honeycomb lattice of ultra-small elements involves the synthesis of diblock template to use as an etching mask, reactive ion etching, chemical processing and near parallel deposition of permalloy material ($Ni_{0.81}Fe_{0.19}$) on top of honeycomb substrate in an electron-beam evaporator in a high vacuum environment (see Supplementary Materials for the schematic description of the fabrication scheme). The atomic force micrograph of a typical honeycomb lattice, fabricated using this method, is shown in Fig. 1a. The preliminary characterizations of artificial magnetic honeycomb lattice of ultra-small elements are performed using electrical and magnetic measurements. Electrical measurement was performed using four-probe method. In Fig. 2, we show the plot of electrical resistivity vs. temperature. Three features become immediately obvious in the experimental data: (a) electrical resistivity depicts a broad peak type feature, centered around $T \simeq 215$ K. (b) The broad peak is followed by a gradual enhancement in resistivity below $T \simeq 135$ K. (c) At much lower temperature, $T \leq 25$ K, the electrical resistivity registers a significant enhancement in a very narrow temperature range, as if it is an insulator. As a result, there are four temperature regimes, as shown by the dashed lines in the figure. The broad peak possibly indicates the onset of a short-range magnetic order in the system. Similar observation in electrical resistivity was previously ascribed to the onset of a short-range magnetic order in a magnetic material[19,20]. The strong enhancement in resistivity at much lower temperature is unprecedented. Interestingly, magnetic measurements of the honeycomb lattice also reveal four somewhat similar regimes, see Fig. S2 in the Supplementary Materials, where net magnetic moment tends to attend the zero magnetization state as $T \rightarrow 0$ K. The four temperature regimes in electrical and magnetic measurements suggest the existence of temperature dependent different magnetic phases in the system. As temperature reduces, the system undergoes a transition from paramagnetic spin gas state to the short-range ordered spin ice state, as manifested by the broad peak in resistance. It is followed by another short-range ordered magnetic charge configuration at further lower temperature. At low enough temperature, T → 0 K, the system tends to attain the spin solid state of zero magnetization state (see Fig. S2). This can be the reason behind the divergence in electrical resistance at very low temperature when the system enters into a strong insulating regime of the spin solid state, consisting of the ordered arrangement of the vortex loops of opposite chiralities. Detailed neutron scattering measurements, as discussed in the following paragraphs, not only corroborate these arguments but also provide new information.

**Small angle neutron scattering measurements on artificial honeycomb lattice of ultra-small elements.** Next, we have performed small angle neutron scattering measurements to gain more insight into the development of various magnetic regimes (see Methods for detail about the measurement procedure)[21]. SANS measurements were performed with incident neutron wavelength of 6 Å at four different temperatures, representative of four different regimes in Fig. 2: 300 K, 175 K, 40 K and 6 K. In Fig. 3, we have plotted the net intensity as a function of reciprocal wave vector, q, at $T = 6\,K$, $40\,K$ and $175\,K$, subtracted by $T = 300\,K$ data. Two sets of peaks with positive intensities are observed at different wave-vectors of $q = 0.032\,Å^{-1}$ and $0.0033\,Å^{-1}$ in 1 m and 13 m detector's positions in Fig. 3a and g, respectively. Additionally, a weak but broad peak with negative intensity is also observed at $q = 0.0102\,Å^{-1}$ in 4 m detector position, suggesting higher intensity at $T = 300\,K$. The best fit to the experimental data is obtained using a Gaussian line-shape, given by $I \propto \exp(-((q-q_0)/\text{width})^2)$. The fitted values of the width of the peaks, centered at $q = 0.032\,Å^{-1}$ and $0.0033\,Å^{-1}$, are found to be $0.0055(20)\,Å^{-1}$ and $0.00052(18)\,Å^{-1}$, respectively. The width of the peak at lower q is comparable to the instrument resolution. Corresponding full width at half maximum (FWHM), estimated using the standard Gaussian description of FWHM = 1.665 × (width of Gaussian curve), are 0.00915 and 0.00086, respectively. FWHM of the peak at higher q is more than an order of magnitude larger than the FWHM at lower q. In real space, the broad peak corresponds to a spatial correlation, $\Delta r = 2\pi/\text{FWHM}$, of $\simeq 70\,nm$, while the resolution limited peak at smaller q suggests a long range order correlation, exceeding over $750\,nm$ in length. In physical term, the correlation length of $\simeq 70\,nm$ corresponds to the size of two honeycombs, basically indicating the coexistence of a short-range order with long range order at $T = 175\,K$. A similar analysis of the negative peak at $q = 0.0102\,Å^{-1}$ yields a $\text{FWHM} = 0.02\,Å^{-1}$, corresponding to the correlation length of $\simeq 30\,nm$. This is larger than the length of a honeycomb element but smaller than the size of a honeycomb. So, the broad peak at $T = 300\,K$ may be arising due to the random spin ice type correlation between moments, as shown in Fig. 1b; also expected in a paramagnetic spin gas state. After all, the spin gas state consists of '2-in & 1-out' (or vice-versa) moment arrangements. As shown in Fig. 3e and f, the broad peak at $q = 0.0102\,Å^{-1}$ becomes more pronounced as temperature decreases. It suggests that the random spin ice type correlation is not completely suppressed at $T = 175\,K$.

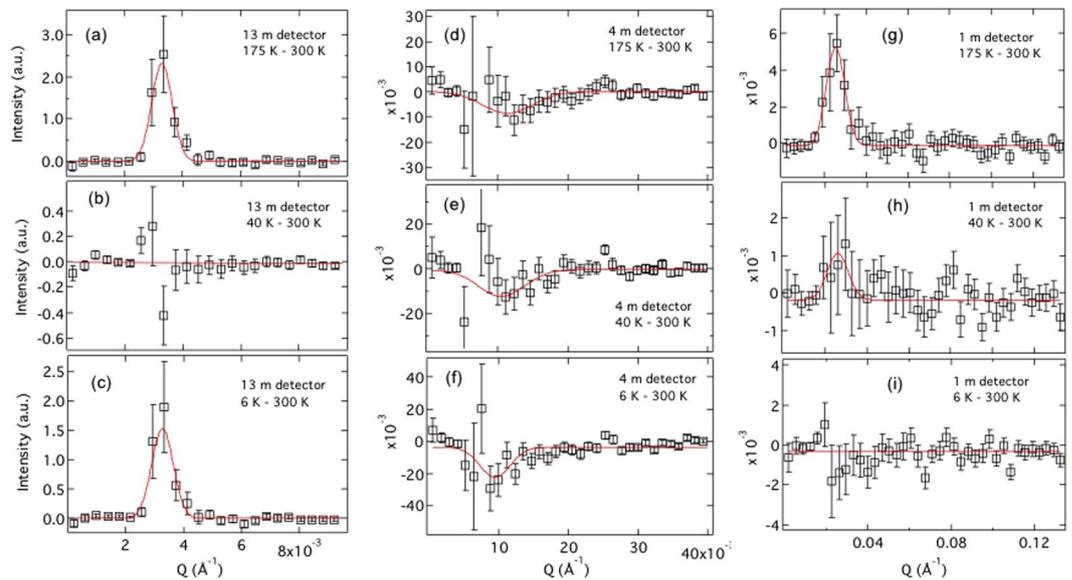

**Figure 3.** SANS measurements at different temperatures. Experimental data were collected in three detector positions of 13 m, 4 m and 1 m. (**a–c**) Plots of intensity vs wave vector, obtained in 13 m detector position. Sharp resolution limited peaks are observed in $T = 175\,K - 300\,K$ and $6\,K - 300\,K$ figures. No such behavior is observed at $T = 40\,K$. Experimental data are well described by Gaussian line-shape. (**d–f**) Results in 4 m detector position. Broad negative peaks (note the q-range) are observed at all three temperatures. The intensity of the negative peaks at $T = 40\,K$ and $6\,K$ are somewhat comparable, but higher than that at $175\,K$. It suggests that the short-range order, which onsets at $T = 300\,K$, is completely suppressed at low temperature. (**g–i**) Plots in 1 m detector position. A significant but broad peak, spanning over the size of two honeycombs (see text), arise when temperature is reduced to $T = 175\,K$. For further reduction in temperature, the peak weakens significantly before disappearing altogether at $T = 6\,K$.

**Reentrant behavior of long range ordered state.** Since an artificial honeycomb is expected to manifest only three magnetic phases (spin ice, charge ordered and spin solid states) with distinct length scales[7,11,12], the estimated lengths give important hint about the nature of underlying magnetism in artificial honeycomb lattice of ultra-small element. The spatial correlations in spin ice, charge ordered and spin solid phases are depicted by spatial correlations across one vertex (less than the size of a honeycomb), two honeycombs due to two vortex loops of opposite chiralities and a long range order, respectively[7]. Accordingly, it can be inferred that the underlying magnetism at $T = 175\,K$ consists of a charge ordered state and a long-range ordered spin solid state in this system. Also, the random spin ice correlation, which is onset at $T = 300\,K$, is not completely suppressed. The system certainly does not manifest unique magnetic configuration, as predicted by recent theoretical researches. It is a combination of all three magnetic phases. When the measurement temperature is reduced to $T = 40\,K$, the peak corresponding to the long range ordered state in 13 m detector position disappears. Only the weak remnants of the peak centered around $q = 0.032\,Å^{-1}$ in 1 m detector position is observed, see Fig. 3h. The width of the peak, as estimated from the Gaussian line-shape, is comparable to that found at $T = 175$ K in Fig. 3g. It suggests that magnetic charge ordered state survives to this temperature, albeit weakly. Comparable negative intensities in Fig. 3e and f, that are also significantly higher than in Fig. 3d, rules out the presence of random spin ice correlation at $T = 40$ K or 6 K.

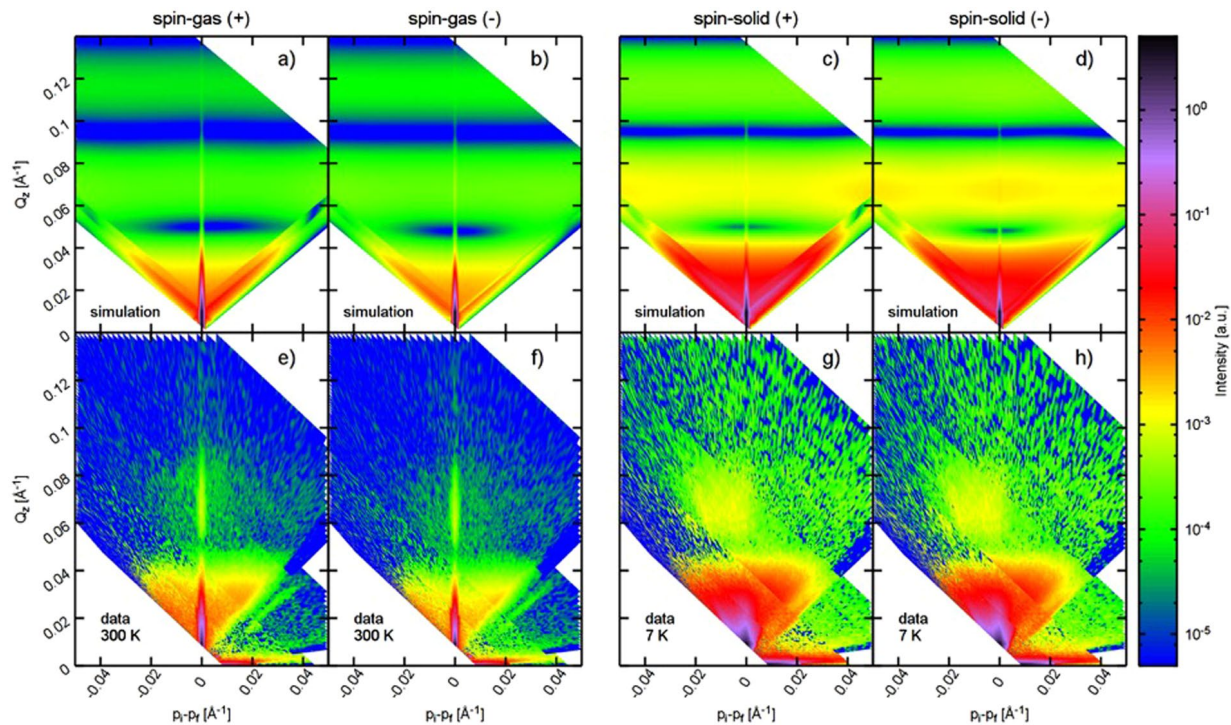

**Figure 4.** Polarized off-specular neutron reflectometry data in spin-up and spin-down channels at $T = 300$ K and 7 K (lower panels, Fig. **e–h**). Vertical line across the origin represents the specular reflectivity. Experimental data (lower panels, Fig. **e-h**) are in good agreement with the numerically simulated profiles (upper panels, Fig. **a–d**) for paramagnetic spin gas at 300 K and the spin solid state at 7 K (as shown in Fig. 1e)

In a surprising observation, SANS measurements at $T = 6$ K reveal the re-entrant characteristic of the long-range ordered state. As shown in Fig. 3c, the sharp instrument-resolution limited peak, centered at $q = 0.0033$ Å$^{-1}$, reappears as measurement temperature is reduced below the temperature corresponding to the inter-elemental energy ($\simeq 15$ K). Interestingly, the peak exhibits similar characteristics, such as the width or the intensity of the Gaussian peak, as found at $T = 175$ K. It indicates that the nature of long-range correlation is similar at both temperatures. No indication of any other magnetic configuration is detected at this temperature. We have further investigated the nature of underlying magnetic correlation behind the re-entrant long range ordered state in this system using polarized reflectometry measutements. An artificial magnetic honeycomb lattice is predicted to manifest only one type of long range ordered state of the spin solid magnetic phase in $T \to 0$ K limit[7,11].

**Confirmation of spin solid long range order using polarized reflectometry results.** The reflectometry measurements allow us to understand the development of the in-plane magnetic structure in the system (see Methods for details about the measurement technique). In Fig. 4e–h, we plot the off-specular data in the spin up and spin-down polarization channels at $T = 300$ K and 7 K, respectively. Here the vertical direction corresponds to the out-of-plane correlation and the horizontal to the in-plane correlation, given by: $Q_z = \frac{2\pi}{\lambda}(\sin \alpha_i + \sin \alpha_f)$ and $p_i - p_f = \frac{2\pi}{\lambda}(\sin \alpha_i - \sin \alpha_f)$ where $\alpha_i$ and $\alpha_f$ are incidence and outgoing angles of neutron beam[22]. The vertical line across the origin represents the specular reflectivity. Clearly, the off-specular scattering plots show remarkable differences between high and low temperature measurements. At $T = 300$ K, the specular intensity is more than two orders of magnitude stronger compared to the off-specular data, which is the expected behavior for most systems. There is also significant intensity in the off-specular regions due to the spin ice short-range order and the honeycomb structure itself. The difference between the spin-up and the spin-down components in the off-specular reflections is similar to that observed in the specular data. On the other hand, measurements at $T = 7$ K reveal significant increase in the off-specular signal (notice the logarithmic color scale). Also, no specular beam can be distinguished from the off-specular background and the difference between the spin-up and the spin-down components vanishes.

As the nuclear structure will not change significantly upon cooling, this can only be explained by a significant change in the magnetic order. The signal itself is very flat along the x-direction, suggesting the development of an in-plane magnetic correlation. Numerical simulation of the scattering profiles, see Fig. 4a–d, using a paramagnetic phase and vortex magnetic configuration of the spin solid state, as shown in Fig. 1e, reproduces essential features of the experimental data at $T = 300$ K and 7 K, respectively. Numerical simulations also show similar bands of broad scattering along the horizontal direction and an almost negligible specular reflection in the spin solid phase. It confirms the previously established notion that the spin solid state is the only magnetic phase with a long range order in artificial honeycomb lattice.

## Discussion

Finally, we summarize the experimental observations and possible implications. The investigations of nanostructured artificial honeycomb lattice of ultra-small element have provided new details about the evolution of magnetic phases as a function of temperature. Contrary to the present understanding that an artificial magnetic honeycomb lattice undergoes through different magnetic phases, each of them unique to a temperature range[11,12], our investigations suggest that all three magnetic states, spin ice, charge ordered and spin solid states, arise when the temperature is reduced (Fig. 1c). In other words, it does not develop unique magnetic state after crossing over from the paramagnetic spin gas state to lower temperature phases. According to recent theoretical researches, the temperature dependent evolution of magnetic phases in artificial honeycomb lattice is controlled by the total entropy of the system. Each state is identified with a finite entropy. This does not seem to be the case here. Rather, the system is manifested by competing magnetic states when temperature is reduced below the spin gas. After all, the three magnetic phases consist of 2-in & 1-out (or vice-versa) local configurations. Therefore, it is not surprising that when temperature is reduced, the local configurations that are distributed across the lattice tend to attend the minimum energy configurations, which may not necessarily be a global minimum. However, at much lower temperature, $T \to 0$ K, the dominance of entropy in dictating the ground state prevails. Although theoretical descriptions are adept in describing the ground state of the system at low enough temperature[23], clearly new formulation is needed to explain the magnetism at intermediate temperature.

## Methods

Small angle neutron scattering experiment was performed on 30 m SANS instrument at NIST Center for Neutron Research. For this purpose, eight samples of nearly $0.6 \times 0.8$ sq. inch area were stacked on top of each other. The stacked samples were mounted at the end of a cold finger of a closed cycle refrigerator with a base temperature of $\simeq 5$ K. SANS measurements were performed with incident neutron wavelength of 6 Å. A cold Be-filter was employed before the sample to clean the beam. Experimental data were collected in three detector configurations of 1 m, 4 m and 13 m. By varying the detector position, we gain access to large Q-range in the reciprocal space. Measurements were repeated at various temperatures between $T = 300$ K and 6 K. For the reflectometry experiment, measurements were performed on a $25 \times 25$ mm² surface area sample at the magnetism reflectometer, beam line 4 A of the Spallation Neutron Source (SNS), at Oak Ridge National Laboratory. The instrument utilizes the time of flight technique in a horizontal scattering geometry with a bandwidth of $\simeq 2.8$ Å (wavelength varying between 2.2−5.0 Å). The polarized measurements allow us to extract both paramagnetic and structural information, from the specular component, and the diffuse band due to the development of new magnetic phase in the off-specular data. The beam was collimated using a set of slits before the sample and measured with a 2D position sensitive ³He detector. The sample was mounted on the cold finger of a close cycle refrigerator with a base temperature of $T = 7$ K. Beam polarization and polarization analysis was performed using reflective super-mirror devices, achieving better than 98% polarization efficiency over the full wavelength band. Electrical measurements were performed using four probe method in closed-cycle refrigerator based cryostat with a base temperature of 5 K. The electrical contacts were made using commercially available silver paint and gold wire.

## References

1. Wang, R. *et al.* Artificial spin ice in a geometrically frustrated lattice of nanoscale ferromagnetic islands. *Nature* **439**, 303–306 (2006).
2. Starykh, O. Unusual ordered phases of highly frustrated magnets: a review. *Rep. Prog. Phys.* **78**, 052502 (2015).
3. Gardner, J., Gingras, M. & Greedan, J. Magnetic pyrochlore oxides. *Rev. Mod. Phys.* **82**, 53 (2010).
4. Bader, D. S. Colloquium: Opportunities in nanomagnetism. *Rev. Mod. Phys.* **78**, 1 (2006).
5. Wolf, S. A. *et al.* Spintronics: a spin-based electronics vision for the future. *Science.* **294**, 1488 (2001).
6. Nisoli, C., Moessner, R. & Schiffer, P. Colloquium: Artificial spin ice: Designing and imaging magnetic frustration. *Rev. Mod. Phys.* **85**, 1473 (2013).
7. Branford, W., Ladak, S., Read, D., Zeissler, K. & Cohen, L. Emerging chirality in artificial spin ice. *Science* **335**, 1597 (2012).
8. Tanaka, M., Saitoh, E., Miyajima, H., Yamaoka, T. & Iye, Y. Magnetic interactions in a ferromagnetic honeycomb nanoscale network. *Phys. Rev. B* **73**, 052411 (2006).
9. Qi, Y., Brintlinger, T. & Cummings, J. Direct observation of the ice rule in an artificial kagome spin ice. *Phys. Rev. B* **77**, 094418 (2008).
10. Mengotti, E., Heyderman, L., Rodriguez, A., Nolting, F. & Hugli, R. Real-space observation of emergent magnetic monopoles and associated Dirac strings in artificial kagome spin ice. *Nature Phys* **7**, 68 (2011).
11. Chern, G., Mellado, P. & Tchernyshyov, O. Two-stage ordering of spins in dipolar spin ice on the kagome lattice. *Phys. Rev. Lett.* **106**, 207202 (2011).
12. Moller, G. & Moessner, R. Magnetic multipole analysis of kagome and artificial spin-ice dipolar arrays. *Phys. Rev. B* **80**, 140409 (R) (2009).
13. Hugli, R. *et al.* Artificial kagome spin ice: dimensional reduction, avalanche control and emergent magnetic monopoles. *Phil. Trans. R. Soc. A* **370**, 5767 (2012).
14. Ladak, S., Read, D., Perkins, G., Cohen, L. & Branford, W. Direct observation of magnetic monopole defects in an artificial spin-ice system. *Nature Physics* **6**, 359 (2010).
15. Shen, Y. *et al.* Dynamics of artificial spin ice: a continuous honeycomb network. *New J. Phys.* **14**, 035022 (2012).
16. Sendetskyi, O. *et al.* Magnetic diffuse scattering in artificial kagome spin ice. *Phys. Rev. B* **93**, 224413 (2016).
17. Anghinolfi, L. *et al.* Thermodynamic phase transitions in a frustrated magnetic metamaterial. *Nature Comm* **6**, 8278 (2015).
18. Park, S., Kim, B., Yavuzcetin, O., Tuominen, M. & Russell, T. Ordering of PS-b-P4VP on Patterned Silicon Surfaces. *ACS Nano* **2**, 1363 (2008).

19. Walker, H. *et al*. Magnetic a*nd electrical properties of dh*cp Np Pd 3 and $(U_{1x} Np_x)Pd_3$. *Phys. Rev. B* **76**, 174437 (2007).
20. Singh, D. K. & Tuominen, M. Realization of artificial Kondo lattices in nanostructured arrays. *Phys. Rev. B* **83**, 014408 (2011).
21. Glinka, C. J. *et al*. The 30 m small-angle neutron scattering instruments at the National Institute of Standards and Technolog. y. *J. App. Cryst* **31**, 430 (1998).
22. Lauter, V., Lauter, H., Glavic, A. and Toperverg, B., *Reference Module in Materials Science and Materials Engineering* **2016**
23. Mellado, P., Petrova, O., Shen, Y. & Tchernyshyov, O. Dynamics of magnetic charges in artificial spin ice. *Phys. Rev. Lett.* **105**, 187206 (2010).

### Acknowledgements

We thank Dr. Valeria Lauter and Dr. Kathryn Krycka for helpful discussion. We acknowledge Dr. Artur Glavic's help with the polarized reflectometry analysis. The research at MU was supported by the U.S. Department of Energy, Office of Basic Energy Sciences under Grant No. DE-SC0014461. A portion of this research used resources at the Spallation Neutron Source, a DOE Office of Science User Facility operated by the Oak Ridge National Laboratory. Neutron scattering research at NIST is supported by the Department of Commerce.

### Author Contributions

B.S. and Y.C. contributed equally to this work. The honeycomb samples were fabricated by B.S. The experiments were carried out by B.S., Y.C., A.D. and D.K.S. Data analysis were performed out by Y.C., B.S. and D.K.S. The paper was written by D.K.S. where everyone contributed.

### Additional Information

**Supplementary information** accompanies this paper at https://doi.org/10.1038/s41598-017-15786-8.

**Competing Interests:** The authors declare that they have no competing interests.

**Publisher's note:** Springer Nature remains neutral with regard to jurisdictional claims in published maps and institutional affiliations.